# Generation and detection of gravitational waves at microwave frequencies by means of a superconducting two-body system*

** Professor in the School of Natural Sciences and in the School of Engineering
University of California, P. O. Box 2039, Merced, CA 95344, U. S. A., e-mail:
rchiao@ucmerced.edu

ABSTRACT: The 2-body system of a superconducting sphere levitated in the magnetic field generated by a persistent current in a superconducting ring, can possibly convert gravitational waves into electromagnetic waves, and vice versa. Faraday's law of induction implies that the time-varying distance between the sphere and the ring caused by the tidal force of an incident gravitational wave induces time-varying electrical currents, which are the source of an electromagnetic wave at the same frequency as the incident gravitational wave. At sufficiently low temperatures, the internal degrees of freedom of the superconductors are frozen out because of the superconducting energy gap, and only external degrees of freedom, which are coupled to the radiation fields, remain. Hence this wave-conversion process is loss-free and therefore efficient, and by time-reversal symmetry, so is the reverse process. A Hertz-like experiment at microwave frequencies should therefore be practical to perform. This would open up observations of the gravitational-wave analog of the Cosmic Microwave Background from the extremely early Big Bang, and also communications directly through the interior of the Earth.

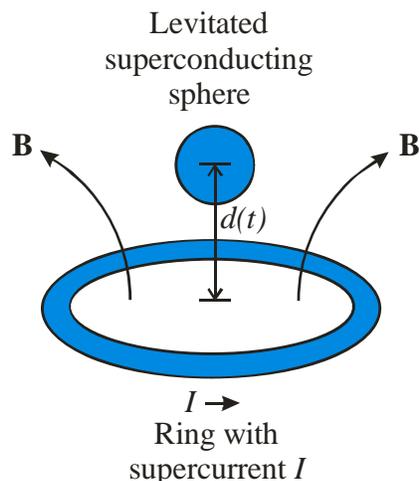

FIGURE 1: The DC magnetic **B** field generated by a superconducting ring with a persistent current $I$ levitates a superconducting sphere. When a (+) polarized gravitational wave propagates radially inwards and converges upon the center of mass of this two-body system, the distance between the sphere and the ring $d(t)$ changes periodically with time, for example, at a microwave frequency, with small-amplitude excursions around an average distance of on the order of the microwave wavelength. This superconducting two-body configuration, viewed as a gravitational-wave antenna, can possibly efficiently convert this incident gravitational (GR) wave into an electromagnetic (EM) wave, and vice versa. Under the operation of time reversal, the current $I$ and the magnetic field **B** of the ring in this Figure are reversed in direction, in order to achieve the time-reversed process of converting the EM wave back into the GR wave.

Consider the configuration of a superconducting sphere levitated above a superconducting ring, as shown in Figure 1 [this was suggested by Clive Rowe, personal communication]. For levitation to occur, it is required that the downwards force of Earth's gravity $\mathbf{F}_{GR}{}^{(0)}$ on the sphere be exactly balanced by the upwards electromagnetic force $\mathbf{F}_{EM}{}^{(0)}$ on the sphere due to the complete expulsion of the **B** field in the Meissner effect [1], so that

$$\mathbf{F}_{EM}{}^{(0)} = -\mathbf{F}_{GR}{}^{(0)}.$$

(The zero superscripts denote forces that are evaluated in the zeroth order of perturbation theory, i.e., in the absence of any perturbing radiation fields.)

Now let a single mode of a weak, linearly polarized gravitational wave [2] at a microwave frequency (such as from the Big Bang), with one of its polarization axes along the vertical axis, be incident upon this two-body system. For simplicity, let us assume that the mass of the ring is much larger than that of the sphere, so that the reduced mass of the two-body system becomes simply that of the sphere, and the center of mass of the system approximately coincides with the center of the ring. The gravitational wave then exerts a time-varying tidal force upon the sphere relative to the ring. According to a distant inertial observer, the space between the sphere and the ring is being periodically squeezed and stretched with time, so that the distance *d(t)* is a sinusoidal function of time.

The mechanism for the wave conversion is the following: In the presence of the DC **B** field of the superconducting ring, Faraday's law of induction implies that the time-varying distance *d(t)* between the sphere and the ring causes induced electrical currents to circulate azimuthally around the bottom pole of the sphere at a microwave frequency, and similarly, induced countercurrents around the ring. These time-varying induced currents are a source of electromagnetic radiation at the same frequency as that of the incident gravitational radiation. Thus a gravitational (GR) wave can in principle be converted into an electromagnetic (EM) wave [3]. Lenz's law implies that there exists an instantaneous EM force $\mathbf{F}_{EM}$ acting on the sphere which *opposes* the instantaneous GR wave tidal force $\mathbf{F}_{GR}$ acting on the sphere, such that under certain circumstances (to be discussed below) quasi-static mechanical equilibrium holds, i.e.,

$$\mathbf{F}_{EM} = -\mathbf{F}_{GR}.$$

Note that this equation implies an equality of the magnitudes of the electromagnetic and gravitational forces on the sphere.

Let us now examine whether or not there can exist circumstances under which a complete conversion of incoming GR wave power into outgoing EM wave power in Figure 1 can in principle occur. Under these circumstances, energy conservation demands that, in terms of quantities as measured by the distant inertial observer,

$$\text{Power absorbed from GR wave } = \langle \mathbf{F}_{GR} \cdot \mathbf{v}_{rad} \rangle = -\langle \mathbf{F}_{EM} \cdot \mathbf{v}_{rad} \rangle = \text{Power emitted into EM wave.}$$

The angular brackets denote the time average over one period of the radiation fields, and $\mathbf{v}_{rad}$ is the instantaneous radiation-damping velocity of the sphere undergoing rigid-body motion relative to the ring, which is parallel to, and in phase with, the instantaneous force $\mathbf{F}_{GR}$, as seen by the distant observer. These circumstances can occur in a regime in which the electromagnetic radiation damping of the motion of the superconducting two-body system is the dominant damping mechanism as compared to all other loss mechanisms.

The superconducting sphere and ring can both undergo rigid-body motion in response to the gravitational wave, because the quantum adiabatic theorem [3,4] applies to all perturbations due to any kind of radiation, whenever the frequency of these perturbations is below the BCS gap frequency [1]. Superconductors stay adiabatically, hence rigidly, in the BCS ground state of the system, because no internal excitations are allowed within the BCS energy gap [1,3,5].

Also, there must be little energy dissipation, such as into heat, other than that due to radiation damping or conversion. This requirement is satisfied by the strictly zero resistance of the superconductors, which follows from the fact that no dissipative excitations into heat due to ohmic processes (or phonon generation to be discussed later) are allowed within the BCS energy gap, when the frequency of the radiation is lower than the BCS gap frequency [1,3,5].

Thus we see that a complete, loss-free conversion from GR wave energy to EM wave energy can indeed occur, provided that all internal degrees of freedom of the superconducting two-body system are frozen at low temperatures, so that only its external degrees of freedom remain. For these external degrees of freedom, in the case when electromagnetic radiation damping is dominant, there then can exist a quasi-static mechanical equilibrium at each instant of time, such that the instantaneous forces obey the equality $\mathbf{F}_{GR} = -\mathbf{F}_{EM}$, just like in the case of a linear induction motor (or generator). Then the power in a single, incoming GR wave mode can be completely converted into the power in a single, outgoing EM wave mode.

Under the operation of time reversal, the single, outgoing EM wave mode becomes a single, incoming EM wave mode incident upon the superconducting two-body system. The EM wave mode is then back-converted into an outgoing GR wave mode. This reverse (or reciprocal) process has the same complete, loss-free conversion efficiency as the forward process by time-reversal symmetry [3,6]. Energy conservation demands that

$$\text{Power absorbed from EM wave} = \langle \mathbf{F}_{EM} \cdot \mathbf{v}_{rad} \rangle = -\langle \mathbf{F}_{GR} \cdot \mathbf{v}_{rad} \rangle = \text{Power emitted into GR wave.}$$

But this is satisfied by Lenz's law and the equality $\mathbf{F}_{GR} = -\mathbf{F}_{EM}$ that is also valid for the reverse process. Here the force $\mathbf{F}_{EM}$ is squeezing and stretching periodically the space between the sphere and the ring, so that the sphere is acting *upon* space, not moving *through* space, in a time-reversal of the forward process.

Under the action of the gravitational wave, the space between the sphere and the ring is undergoing small amplitude, anisotropic, periodic time-varying strains $h_{ij}$ of the metric tensor [2], as if the space were an elastic medium. According to the Equivalence Principle [2], in both the forward and the reverse processes, the sphere is not accelerating at all with respect to a tiny, local inertial observer located at its center, so that the sphere is not moving, i.e., accelerating, *through* space. Hence the objection that its inertia limits the amount of emitted GR radiation to extremely small values is invalid; see section 12 of [3a]. Similarly, the objection that the motion of the sphere through a flat-space background must be relativistic (i.e., with velocities close to *c*) in order to generate any appreciable amount of GR waves is also invalid.

Therefore we propose here to perform a Hertz-like experiment at 12 GHz, in which the EM-to-GR wave-conversion process becomes the *source* of GR waves, and the GR-to-EM wave-conversion process becomes the *receiver* of GR waves. Faraday cages consisting of normal metals at room temperature (the metallic casings of the two well-separated liquid helium storage dewars to be described below) prevent the transmission of EM waves, so that only GR waves, which can easily pass through all ordinary, classical matter, such as the normal (i.e., dissipative) metals of which standard, room-temperature Faraday cages are composed, are transmitted between the two halves of the apparatus that serve as the source and the receiver, respectively [3]. Such an experiment should be practical to perform using standard microwave sources and receivers, since the scattering cross-sections and the wave conversion efficiencies of the two-body superconducting system such as the one depicted in Figure 1, when viewed as a gravitational-wave antenna, should be large enough to be experimentally interesting [3; see also below].

We plan to use lead (Pb) as the type I superconducting material for both the sphere and the ring. The sphere could consist of a light, low-density material (e.g., Styrofoam) coated on its surface with a Pb film which is much thicker than the penetration depth [5]. The ring could consist of a flat copper gasket electroplated with a thick coating of Pb. The central hole of the gasket could have a diameter of 34 mm, which should levitate the superconducting sphere above the hole of the ring at a distance of around 6 mm, i.e., approximately a quarter of a microwave wavelength at 12 GHz where we plan to work (12 GHz is over an order of magnitude below the BCS gap frequency of Pb, so that the losses at this frequency are negligible [3]). A persistent current in the ring sufficient to levitate the sphere could be induced by means of a nearby permanent magnet, which would then be removed after the ring has been cooled to become superconducting below the transition temperature of Pb at 7.2 K [5]. We plan also to experiment with other two-body geometries, such as with a pair of superconducting coaxial rings with opposing trapped magnetic flux [suggested by S. Minter], as realizations of the superconducting two-body system. We have chosen to use a type I superconductor instead of a type II superconductor, in order to avoid microwave losses arising from the Abrikosov vortex-motion degrees of freedom.

We propose to perform the first experiments in a standard 4 K liquid helium storage dewar using a dewar insert, which consists of a 50 mm diameter cylindrical structure with a detachable sample holder that is capable of containing a specific sphere-ring configuration (or other superconducting two-body geometries) at its bottom end. Such a dewar insert could be inserted directly into the liquid helium inside a storage dewar, in order to explore the parameters for the levitation of the sphere by the ring, by adjusting the dimensions of the sphere-ring configuration. Optical access and viewing could be made possible by means of glass prisms placed inside the sample can near the ring. Alternatively, a Hall probe sensor could detect the onset of the Meissner effect and levitation of the sphere. These dewar-insert experiments would enable us to test the computer simulations we are presently conducting for the levitation of type I superconducting spheres by various magnetic field configurations.

Two of these dewar inserts, each with a sphere-ring configuration (or some other superconducting two-body configuration) immersed deep inside two well separated 4 K liquid helium dewars, with the associated microwave electronics, would then allow us to perform the Hertz-like experiment. In preparation for this experiment, we plan to determine at room temperature the microwave antenna design parameters of the sphere-ring and the two-coaxial-ring configurations (and other possible configurations) using a network analyzer. We plan to use an appropriately designed magnetic loop antenna to couple to the $TE_{01}$ microwave cylindrical mode of a superconducting ring for transferring energy from the GR wave mode to the EM wave mode, and vice versa.

We have already built and tested microwave sources and receivers at 12 GHz [6]. We have used a frequency-doubled 6 GHz microwave-cavity oscillator as the source, with an output power level at 12 GHz of around 10 mW, and have measured that a standard, commercial satellite-dish KU band receiver [Precision model PMJ-LNB KU gold label series from Astrotel Communications Corp.] has a noise figure of 0.6 dB at 12 GHz, so that our receiver system Noise-Equivalent-Power (NEP) sensitivity will be on the order of $10^{-25}$ Watts/Hz$^{1/2}$. This will allow us to achieve a very good signal-to-noise ratio for the Hertz-like experiment [3], assuming that our estimates for the cross-sections and wave-conversion efficiencies for the sphere-ring system are of the correct orders of magnitude.

If we should be successful in the Hertz-like experiment, one of the first tests to see if we truly have GR rather than EM coupling between the transmitter (or source) and receiver (or detector) halves of the apparatus, is to tilt the transmitter part of the apparatus by $+22.5^o$ with respect to the vertical around the line of sight joining the transmitter and receiver, and to tilt the receiver part of the apparatus by $-22.5^o$ with respect to the line of sight. The signal should be extinguished at the resulting $45^o$ relative orientation between the two halves of the apparatus, and not at $90^o$, as would the case for EM waves. This would be a clear signature that we have successfully generated and detected GR waves rather than EM waves [this was suggested by Kirk Wegter-McNelly, personal communication].

Another of our initial goals if we should be successful in detecting a signal in the Hertz-like experiment would be to directly measure the speed of gravitational waves for the first time. This can be done either by changing the distance between the two experimental dewars containing the two sphere-ring (or other) configurations, and then measuring the resulting change in phase of the received signal relative to that of the source signal, or by modulating the carrier signal, and measuring the delay in the arrival time of the demodulated signal. We will thereby be able to directly check experimentally in the laboratory for the first time the theoretical prediction by Einstein that the speed of these waves is identical to that of light in vacuum $c$ = 3.00 x $10^8$ meters per second [2] , to within $\pm 1\%$ accuracy.

It may be objected that sound waves will be generated as a dissipative mechanism in the proposed experiment. However, sound waves have a speed that is typically five orders of magnitude smaller than *c*. If Einstein were correct, then the typical generated sound wavelength of 250 nm or smaller would be much less than the gravitational wavelength of 25 mm, so that the typical sound wave would be badly mismatched to the incident GR wave. The generation of sound waves (or phonons) within the superconductor at microwave frequencies inside the BCS gap by the incident gravitational wave would therefore be forbidden due to the Mossbauer-like (or zero-phonon) effect described in section 9 of [3a]; see also [3b]. Hence the sphere and the ring would undergo rigid-body motion in response to the incident gravitational wave.

It may also be objected that the microwave-frequency electrical currents induced through Faraday's law in the GR to EM wave conversion process may be too feeble and too difficult to detect. However, we are not proposing to detect these *currents*, but rather to detect the *power* converted from the GR wave into the EM wave, and vice versa. With an antenna design which takes into account impedance-matching and cross-section considerations properly, the power transmitted from the source to the receiver should be readily measurable.

To better understand the cross-section estimate for the above proposed experiment, it is useful to introduce a weak-field representation of the linearized Einstein equations for the gravito-electric field $\mathbf{E}_G \equiv \mathbf{g}$ (i.e., the gravitational acceleration of a test mass), and for the gravito-magnetic field $\mathbf{B}_G$ (i.e., the far-field, time-varying analog of the Lense-Thirring field), as observed in the coordinate system of a distant inertial observer. This representation takes the form of the Maxwell-like equations [7]

$$\nabla \cdot \mathbf{E}_G = -\frac{\rho_G}{\varepsilon_G} \tag{1.1}$$

$$\nabla \times \mathbf{E}_G = -\frac{\partial \mathbf{B}_G}{\partial t} \tag{1.2}$$

$$\nabla \cdot \mathbf{B}_G = 0 \tag{1.3}$$

$$\nabla \times \mathbf{B}_G = \mu_G \left( -\mathbf{j}_G + \varepsilon_G \frac{\partial \mathbf{E}_G}{\partial t} \right) \tag{1.4}$$

where $\rho_G$ is the mass density and $\mathbf{j}_G$ is the mass current density of slowly-moving matter in the source, and the gravitational analog of the electric permittivity of free space $\varepsilon_0$ is

$$\varepsilon_G = \frac{1}{4\pi G} = 1.19 \times 10^9 \text{ SI units}, \tag{1.5}$$

where $G$ is Newton's constant, and the gravitational analog of the magnetic permeability of free space $\mu_0$ is

$$\mu_G = \frac{4\pi G}{c^2} = 9.31 \times 10^{-27} \text{ SI units}, \tag{1.6}$$

where $c$ is the speed of light.

The fields $\mathbf{E}_G \equiv \mathbf{g}$ and $\mathbf{B}_G$ are measurable quantities that obey the gravitational analog of the Lorentz force law [7]

$$\mathbf{F} = m \left( \mathbf{E}_G + 4\mathbf{v} \times \mathbf{B}_G \right), \tag{1.7}$$

where $m$ is the mass of a test particle, and $\mathbf{v}$ is its velocity, as seen by the distant inertial observer.

These Maxwell-like equations are *linear*, so that the fields obey the superposition principle not only in the vacuum outside of the source, but also in the matter inside the source, provided that the field strengths are sufficiently weak and the matter is sufficiently slowly moving, so that there exists a regime of a *linear response* of the matter to the applied fields. The resulting optics for gravitational waves is therefore *linear*, just like the linear optics for electromagnetic waves.

Therefore any argument involving the characteristic power scale [2]

$$P_o = \frac{c^5}{G} = 3.6 \times 10^{52} \text{ Watts} \qquad (1.8)$$

is irrelevant in the context of linear optical devices such as linear mirrors, since it should always be possible to reduce the intensity of the incident GR waves so that there exists a linear regime of response of such devices to these sufficiently weak GR wave amplitudes. Then there cannot exist any characteristic scale of power in any *linear* response of such devices, such as that given by Equation (1.8).

Also, it is important to emphasize that the matter generating the GW waves in such linear devices as the two-body superconducting system described above, needs not be moving relativistically with velocities close to the speed of light *c* relative to each other, in order to generate any appreciable amount of GR waves. By the time-reversal symmetry argument given above, it is clear that the generation of GR waves by the two-body superconducting system is not due to the matter moving *through* space, but rather is due to the matter acting *upon* space directly. Hence it is irrelevant whether such matter is moving close to *c* through space or not.

In the case of the vacuum, where $\rho_G$ and $\mathbf{j}_G$ both vanish, these equations lead to wave propagation at a wave speed exactly equal to the speed of light

$$c = \frac{1}{\sqrt{\varepsilon_G \mu_G}} = 3.00 \times 10^8 \text{ SI units}. \qquad (1.9)$$

A plane wave solution of the Maxwell-like equations possesses the gravitational characteristic impedance of free space given by [8,3]

$$Z_G = \sqrt{\frac{\mu_G}{\varepsilon_G}} = 2.79 \times 10^{-18} \text{ SI units}, \qquad (1.10)$$

which is the analog of the electromagnetic characteristic impedance of free space

$$Z_0 = \sqrt{\frac{\mu_0}{\varepsilon_0}} = 377 \text{ ohms}. \qquad (1.11)$$

The gravitational characteristic impedance of free space $Z_G$, like $Z_0$, plays a central role in all radiation problems, such as in a comparison of the radiation resistance of gravitational-wave antennas to the value of this impedance, in order to estimate the coupling efficiency of these antennas to free space. The numerical value of $Z_G$ is extremely small, but the impedance of all material objects must be "impedance matched" to this extremely small quantity before significant power can be transferred efficiently from gravitational waves to these objects, or vice versa.

In contrast to the electromagnetic-wave case, in the gravitational-wave case, all ordinary, classical matter, such as Weber bars, possesses impedances much larger than that of the gravitational characteristic impedance of free space $Z_G$. It is therefore extremely difficult to impedance-match gravitational waves to any ordinary, classical matter. As a consequence, it is a general rule that all ordinary, classical matter, such as a Weber bar, is essentially completely transparent to these waves.

However, phase-coherent, loss-free quantum matter, such as superconductors, which can possess strictly *zero* dissipation due to the presence of the BCS energy gap, can be exceptions to this general rule. Experimental evidence for this "quantum dissipationlessness" is the fact that persistent currents in annular superconducting rings have been observed to last much longer than the age of the Universe [1]. One important consequence of the zero-dissipation property of a superconductor is that a mirror-like reflection of sufficiently low-frequency gravitational waves can occur at a planar interface between the vacuum and the superconductor. (By "low frequency" we mean frequencies much less than the BCS gap frequency.)

In the electromagnetic case, the reflection coefficient R of a wave being transmitted down a transmission line with impedance $Z$, which is terminated by a resistance $R$ which vanishes like that of a superconductor, is given by

$$R = \left|\frac{Z-R}{Z+R}\right|^2 \to 100\% \text{ as } R \to 0. \qquad (1.12)$$

This implies that a mirror-like reflection of the wave occurs from a superconductor, when it is used as a short-circuit termination of the transmission line.

Similarly, the reflection coefficient $R_G$ of a sufficiently low-frequency gravitational wave from a superconductor-vacuum interface is given by

$$R_G = \left|\frac{Z_G-R_G}{Z_G+R_G}\right|^2 \to 100\% \text{ as } R_G \to 0, \qquad (1.13)$$

where the gravitational characteristic impedance of free space $Z_G$ is given by Equation (1.10), and $R_G$ is the gravitational analog of the resistance $R$ of the superconductor. This implies that a mirror-like reflection of a sufficiently low-frequency gravitational wave could in principle occur from the surface of the superconductor.

Therefore mirrors for gravitational waves can in principle exist. Curved, parabolic mirrors can focus these waves, and Newtonian telescopes for gravitational radiation can in principle be constructed. In the case of scattering of gravitational waves from superconducting bodies, the above mirror-like-reflection condition implies hard-wall boundary conditions at the surfaces of these superconducting bodies, so that the scattering cross-section of these waves from large superconducting bodies can in principle be geometric, e.g., hard-sphere, in size. For example, for a superconductor in the form of a sphere of radius $a$ which is much larger than the wavelength, the cross-section for GR wave scattering by the sphere is given by

$$\sigma_{\text{hard sphere}} = 2\pi a^2, \qquad (1.14)$$

where $a$ is the radius of the superconductor.

It may be objected that any kind of reflection of gravitational waves from matter, including any kind of partial reflection, would constitute a kind of "anti-gravity" effect. Now it is certainly true that only *positive* masses exist in nature, and therefore that the screening of longitudinal *gravito-electric* fields, like the Earth's gravitational field, including its partial screening by any kind of matter, including superconductors as has been falsely claimed in the so-called "Podkletnov effect", is impossible. However, both *positive* and *negative* mass *currents* can exist in nature, and therefore the screening of transverse *gravito-magnetic* fields, like those of a gravitational wave, including the partial screening by the *reflection* of these waves from matter, should indeed be possible. The strength of the reflection can in principle depend on the details of the nature of the matter.

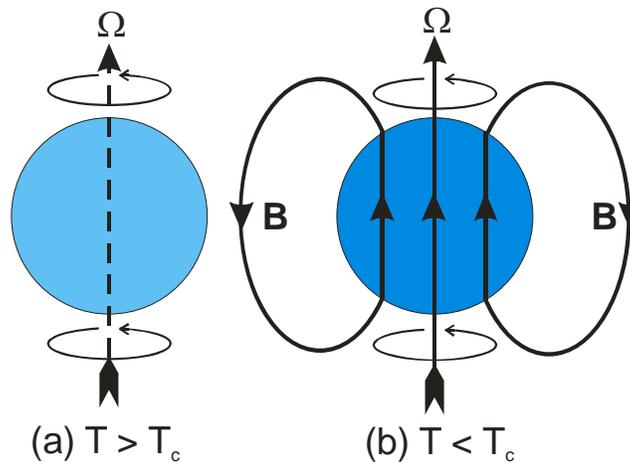

FIGURE 2: Comparison of a rotating normal metal with a rotating superconductor. In (a), the normal metal is a neutral sphere consisting of a rotating superconducting metal at a temperature above its transition temperature $T_c$. In (b), this body is slowly cooled down below $T_c$, whilst it is rotating. Now there is the onset of a magnetic field **B** proportional to its rate of rotation $\Omega$ with respect to the distant "fixed stars".

A closely related objection is that, according to one interpretation of the Equivalence Principle, the response of all kinds of matter, whether classical or quantum mechanical in nature, should be universally the same to all gravitational fields. Locally, all dropped objects undergo exactly the same free-fall motion, which is independent of their mass, composition, or thermodynamic state. Specifically, the response of all matter to gravitational fields, such as the Earth's, is independent of whether the thermodynamic state of the matter is classical or quantum mechanical in nature.

Thus this interpretation of the Equivalence Principle would forbid any difference between the linear response of *incoherent* classical matter, such as Weber bars, and the linear response of *coherent* quantum mechanical matter, such as superconductors, to *any* kind of gravitational field. However, whilst this interpretation is valid for the *local* response of any kind of matter, whether coherent or incoherent, to all *gravito-electric* fields, like that of all dropped, freely-falling objects in their response to the gravitational field of the Earth, it is not valid for the *nonlocal* response of matter, whether coherent or incoherent, to *gravito-magnetic* fields, such as to the Lense-Thirring field arising from the distant matter of the Universe (i.e., from the distant "fixed stars").

Empirical evidence for this latter fact lies in the difference between the response of normal metals and the response of superconductors which are rotating with respect to the distant "fixed stars" [9]. Consider an experiment in which there is a spherical body rotating at a fixed angular speed $\Omega$, which consists of a normal metal that is a superconductor above its transition temperature $T_c$. For all temperatures $T > T_c$, no magnetic field is produced by this neutral, rotating body.

Now consider what happens when this rotating body is slowly cooled, whilst it is undergoing rotation with respect to the distant "fixed stars", down below its transition temperature $T_c$. Now for all temperatures $T < T_c$, there exists a magnetic field **B** produced by this rotating superconducting body in the "London moment" effect. The London moment arises from the constructive quantum interference of the Cooper pairs of electrons in the superconductor near its surface after one round trip around the perimeter of the body, as seen by the distant inertial observer. This quantum interference effect implies that the Cooper pairs everywhere near the surface of the body in its lowest energy state will come to a complete halt with respect to the distant "fixed stars". The nuclei near the surface, however, continue to rotate with respect to the distant "fixed stars". Thus this differential motion of the Cooper pairs with respect to the nuclei near the surface produces a surface electrical current. It is this current that produces the magnetic field **B** which is sketched in Figure 2 (b).

However, according to the above interpretation of the Equivalence Principle, there cannot be *any* difference in the response of the rotating body above or below its transition temperature to *any* kind of gravitational field, including the Lense-Thirring field which arises from the distant "fixed stars". Specifically, the response of the rotating body to the Lense-Thirring field arising from these distant "fixed stars" should be independent of the thermodynamic state of this body, and in particular, it should be independent of whether the rotating body is composed of *incoherent*, classical matter, or *coherent*, quantum mechanical matter. This interpretation of the Equivalence Principle is contradicted by experiments that demonstrate the existence of the London moment [9].

Let us now generalize the above considerations to the case of the time-varying, transverse fields of gravitational radiation, when the frequency of the radiation is much lower than the BCS gap frequency, so that the quantum adiabatic theorem holds. Let us first consider the case of a normal metal, which consists of a superconducting metal above its transition temperature $T_c$. Let this neutral metallic body have the shape of a planar slab with a thickness of half a wavelength of the incident gravitational plane wave, which is propagating towards the slab at normal incidence along the *x* axis from the left, as depicted in Figure 3. The entrance face of the slab is at *x*=0, and the exit face of the slab is at $x=\lambda/2$.

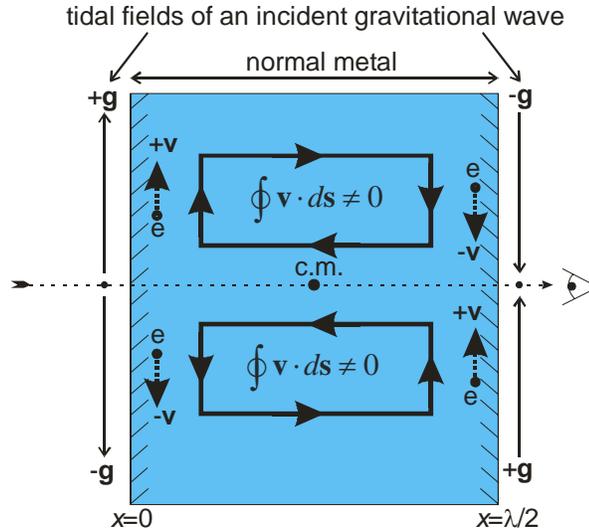

FIGURE 3: A slab of normal metal half a wavelength thick, which is a superconducting metal above its transition temperature $T_c$, in its linear response to the tidal fields of a normally incident gravitational plane wave. The center of mass of the slab is represented by the dot below the label "c.m." The eye on the right represents a distant inertial observer. However, below $T_c$, the upper and lower electrons form a Cooper pair in a Bohm singlet state, i.e., an *entangled* state which violates Bell's theorem and is *nonlocal*. The electrons' *local* free-fall motion is thus suppressed.

The linear response of a normal electron above the superconducting transition temperature $T_c$ near the surface of the metal to the tidal field **g** of the incident wave is represented by the local, instantaneous velocity vector **v**, which would be the free-fall velocity of the normal electron, assuming the absence of any dissipation in the metal. (All measurable quantities are those that are being measured by a distant inertial observer represented by the eye in Figure 3). Since the normal electrons undergo local free fall together with the nearby nuclei (neglecting for the moment the weak restoring forces arising from the ionic lattice), no electrical currents are produced in the linear response of this *neutral* body to the local, instantaneous accelerations due to gravity **g** arising from tidal forces of the gravitational wave.

Let us now choose two rectangular loops, one above the center of mass (c.m.) of the system, and one below it, as shown in Figure 3, in order to evaluate the circulation of the normal electrons around these loops in their linear response to the incident wave. It is obvious upon inspection of Figure 3 that these two closed loop integrals of the velocity **v** of the normal electron, as seen by the distant inertial observer, obey the inequality

$$\oint \mathbf{v} \cdot d\mathbf{s} \neq 0 , \qquad (1.15)$$

so that the circulation of the normal electrons around these loops does not vanish. This is because there is a reversal of the sign of the tidal forces as the wave propagates from the entrance face of the slab at $x=0$ to the exit face at $x=\lambda/2$, and therefore there is also a reversal of the sign of the normal electron velocity in their linear response to these tidal forces.

Next, let us slowly cool this system down below the superconducting transition temperature $T_c$ whilst the incident gravitational wave is still present. The normal electrons now become Cooper pairs, which possess quantum phase coherence over long distances. If the Cooper pairs near the surface of the metal in Figure 3 were to undergo free fall with exactly the same instantaneous, local velocity **v** in their linear response to the tidal fields **g** of the incident gravitational wave, as the instantaneous, local free-fall velocity **v** of the normal electrons in the normal state of the metal (as would be demanded by the above erroneous interpretation of the Equivalence Principle), then their macroscopic wavefunction (or the complex order parameter of Ginzburg and Landau) would no longer by single valued after one round trip around the closed loop, and the constructive quantum interference for these Cooper pairs after a round trip would be impossible. This is because the instantaneous velocity **v** is related to the phase $\phi$ of the Cooper pair macroscopic wavefunction (or of the complex order parameter) by

$$\mathbf{v} = \frac{\hbar}{m}\nabla\phi, \qquad (1.16)$$

where $\hbar$ is Planck's constant/$2\pi$ and $m$ is the mass of a Cooper pair.

However, experiments verifying the existence of the London moment [9] demonstrate that the Cooper pairs do in fact always possess a *single-valued* macroscopic wavefunction (or complex order parameter), as determined by the observations in the inertial frame of the distant observer, i.e., that of the distant "fixed stars". Hence it must be the case that below the temperature $T_c$, the circulation of the Cooper pairs vanishes identically, i.e.,

$$\oint \mathbf{v}\cdot d\mathbf{s} = 0, \qquad (1.17)$$

in the case of the lowest energy state of the Cooper pairs. This implies that their local, instantaneous velocity vanishes everywhere along the left and the right vertical segments at $x=0$ and $x=\lambda/2$ of the arbitrarily chosen closed loops in Figure 3, so that in the limit of arbitrarily skinny loops, it follows that

$$\mathbf{v} = 0 \qquad (1.18)$$

*everywhere* near the surface of the superconductor. Therefore it follows that the time-dependent, *transverse* gravito-electric field

$$\mathbf{E}_G \equiv \mathbf{g} = \left(\frac{\partial \mathbf{v}}{\partial t}\right)_{\text{near surface}} = 0 \qquad (1.19)$$

also vanishes *everywhere* near the surface of the superconductor. This also implies hard-wall boundary conditions that lead to a mirror-like reflection of gravitational waves. (See also footnotes 11 and 12 of [3a].)

The above conclusions also follow from the fact that the entire superconducting slab stays adiabatically in the BCS ground state of the system with a single value of the global phase $\phi$ of the macroscopic wavefunction (or of the complex order parameter) inside the entire superconducting body, i.e.,

$$\phi = \text{constant everywhere} . \qquad (1.20)$$

This is true even in the presence of the weak, local perturbations due to the incident gravitational wave, provided that the wave frequency is well below the BCS gap frequency, so that the wave cannot cause any quantum transitions out of the BCS ground state.

The Cooper pairs in the BCS ground state are in *entangled states* of momentum and spin, which are *nonlocal*, so it is impossible to know whether it is the upper or the lower electron in Figure 3 that is moving upwards or downwards in response to the tidal **g** fields. The large number of these phase-coherent electrons in the system (on the order of Avogadro's number) in the case of superconductors, compensates for the usual weakness of the interaction between gravitational radiation and matter, and leads to a superradiant, mirror-like reflection [3b].

Therefore, just as in the case of the London moment below the transition temperature $T_c$, the Cooper pairs everywhere near the entrance and exit faces of the superconducting slab must come to a complete halt with respect to the distant inertial observer (i.e., with respect to the distant "fixed stars") below $T_c$. The vanishing of **v** everywhere near the surfaces of the slab must be independent of the size of the small amplitude of a sufficiently weak gravitational wave, in the regime of the *linear response* of the superconductor to the wave. This is the behavior of an extremely rigid material in its linear response to the gravitational wave, so that, once again, one concludes that a mirror-like reflection of a sufficiently weak incident gravitational wave should occur at the planar superconductor-vacuum interface. The resulting scattering cross-section for large superconducting bodies should therefore once again be of the order of magnitude of the geometric cross-section given by Equation (1.14). These counter-intuitive predictions will be tested in the above proposed experiment.

It is to be emphasized that throughout the above discussion, all measurable quantities in the above equations are those which are being observed and measured by the distant inertial observer. The coordinate system being used in these measurements is the one set up by means of light signals sent to and from this distant observer, and the time coordinate being used is the one set up by means of this observer's clock [11].

This experiment could lead to important applications in science and engineering. In science, it would open up the possibility of gravitational-wave astronomy at microwave frequencies. One important problem to explore would be observations of the analog of the Cosmic Microwave Background (CMB) in gravitational radiation. Since the Universe is much more transparent to gravitational waves than to electromagnetic waves, such observations would allow a much more penetrating look into the extremely early Big Bang towards the Planck scale of time, than the presently well-studied CMB. Different cosmological models of the very early Universe give widely differing predictions of the spectrum of this penetrating radiation, so that by measurements of the spectrum, one could tell which model, if any, is close to the truth [10]. The anisotropy in this radiation would also be very important to observe.

In engineering, it could open up the possibility of intercontinental communications by means of microwave-frequency gravitational waves directly through the interior of the Earth, which is transparent to such waves. This would eliminate the need of communications satellites, and would allow an economical means of communication with people deep underground or underwater in submarines in the oceans. Such a new direction of gravitational-wave engineering could aptly be called "gravity radio" [3].

Acknowledgments: This work was supported in part by the CTNS-STARS program. I would like to thank Prof. Thomas Kibble for his helpful comments.